\RequirePackage[hyphens]{url}
\documentclass[%
 reprint,
nofootinbib,
 amsmath,amssymb,
 aps,
english,aps,prper,reprint,showpacs,titlepage,longbibliography,nofootinbib,superscriptaddress,floatfix
]{revtex4-2} 

\usepackage{ragged2e}
\usepackage[tight]{subfigure}
\usepackage{graphicx}
\usepackage{dcolumn}
\usepackage{bm}
\PassOptionsToPackage{hyphens}{url}\usepackage{hyperref}

\usepackage{float}

\usepackage{comment}
\hypersetup{colorlinks=true,urlcolor=blue,citecolor=blue,linkcolor=blue}   
\usepackage[T1]{fontenc}	
\usepackage[latin9]{inputenc}	
\usepackage{geometry}		
\usepackage{graphicx}
\usepackage[above,below]{placeins}	
\usepackage{times}
\usepackage{multirow}
\usepackage{ifthen}
\usepackage{makecell}
\usepackage{longtable}

\usepackage{outlines}

\newboolean{redactswitch}
\setboolean{redactswitch}{false} 

\newcommand{\redact}[1]{\ifthenelse{
    \boolean{redactswitch}}{{[Redacted]}}{
    {#1}}}
    
\newenvironment{iquote}
    {\vspace{-.33\baselineskip}\itshape\list{}{\leftmargin=0.15in\rightmargin=0.15in}%
    \item\relax}
    {\endlist\vspace{-.33\baselineskip}}

\newcommand{\fracroot}[2]{\ifthenelse{#1=1}{\frac{1}{\sqrt{#2}}}{\sqrt{\frac{#1}{#2}}}}

\newcommand{\s}[0]{\textasteriskcentered{}}

\usepackage{makecell}

\newif\ifshowtimestamp
\showtimestampfalse

\usepackage{footmisc}

\newif\ifinappendix
\let\oldappendix\appendix
\renewcommand{\appendix}{
  \oldappendix
  \inappendixtrue
}

\begin{document}

\title{Disparities in access to US quantum information education}

\author{Josephine C. Meyer}
 \email{josephine.meyer@colorado.edu}
 \affiliation{Dept.\ of Physics, University of Colorado Boulder, Boulder, CO 80309, USA}
\author{Gina Passante}%
\affiliation{%
 Dept.\ of Physics, California State University Fullerton, Fullerton, CA 92831, USA
}%

\author{Bethany Wilcox}

\affiliation{Dept.\ of Physics, University of Colorado Boulder, Boulder, CO 80309, USA}%

\date{\today}

\begin{abstract}
Driven in large part by the National Quantum Initiative Act of 2018, quantum information science (QIS) coursework and degree programs are rapidly spreading across US institutions. Yet prior work suggests that access to quantum workforce education is unequally distributed, disproportionately benefiting students at private research-focused institutions whose student bodies are unrepresentative of US higher education as a whole. We use regression analysis to analyze the distribution of QIS coursework across 456 institutions of higher learning as of fall 2022, identifying statistically significant disparities across institutions in particular along the axes of institution classification, funding, and geographic distribution suggesting today's QIS education programs are largely failing to reach low-income and rural students. We also conduct a brief analysis of the distribution of emerging dedicated QIS degree programs, discovering much the same trends. We conclude with a discussion of implications for educators, policymakers, and education researchers including specific policy recommendations to direct investments in QIS education to schools serving low-income and rural students, leverage existing grassroots diversity and inclusion initiatives that have arisen within the quantum community, and update and modernize procedures for collecting QIS educational data to better track these trends.
\end{abstract}

\maketitle


\section{Introduction and background}

Quantum information science (QIS) education programs have been proliferating at US institutions in response to urgent calls from the community to support the development of a quantum-ready workforce \cite{Fox:2020,Marrongelle:2020,Plunkett:2020,Aiello:2021,Perron:2021,Hasanovic:2022,Hughes:2022,Singh:2022,Greinert:2023} and an infusion of funding from the National Quantum Initiative Act of 2018 \cite{NQIA:2018}. Such programs range from stand-alone courses \cite{Cervantes:2021} to undergraduate and graduate degree programs \cite{Asfaw:2022, Dzurak:2022, Wong:2022} to dedicated programs targeting high school students \cite{QBQ, Q12}.

However, absent careful attention to equity, there is a risk that quantum education initiatives will end up inadvertently replicating the diversity and access gaps that have long plagued physics and engineering education \cite{Aiello:2021, Singh:2021} -- disparities that are well-documented throughout the physics education research (PER) literature and beyond (e.g.\ \cite{Eddy:2016,Miriti:2020,TEAMUP:2020}). Ten Holter \textit{et al.} argue that an ethical and responsible quantum revolution requires careful attention to equitable distribution of the benefits of quantum technologies \cite{TenHolter:2022}, including but not limited to the educational opportunities and well-paying jobs quantum technologies will enable. Moreover, recent work has established the importance of a quantum-literate society even among those outside the industry \cite{EPSRC:2018,Nita:2021}, lest elite capture of public dialog and policymaking result in the benefits of quantum technologies flowing primarily to the most wealthy and powerful \cite{DiVincenzo:2017}. Equitable access to quantum technologies has also been framed through the lenses of democratization \cite{Seskir:2023} and human rights \cite{Krishnamurthy:2022}. Horace Mann famously called education ``the great equalizer of conditions of men'' \cite{Mann:1848}, but schooling can just as easily serve to restrict as to democratize access to knowledge and resources. We argue that efforts to ensure equitable distribution of the benefits of the quantum revolution must pay particular attention to the education system that feeds into quantum jobs.

A prior study conducted in 2020 found that QIS coursework at US universities was primarily concentrated at large, private R1 research universities, alongside indications of disparities in geographic representation and minority-serving institution (MSI) status \cite{Cervantes:2021}. However, small sample size at the time of the study precluded making quantitative claims about the distribution of QIS education programs and from examining the intersections of multiple institutional factors. Perron \textit{et al.} likewise noted a tendency for undergraduate QIS education programs to be concentrated at large and well-funded doctoral-granting institutions \cite{Perron:2021}. Our work here builds on these observations using quantitative methodologies to quantify the existence and extent of such potential disparities.

\section{Aims and scope}
\label{sec:aims-scope}
Voices in the quantum community are increasingly calling for equity to be made a key value in the development of quantum education programs and quantum workforce development \cite{Aiello:2021,Josten:2022,Wolbring:2022,Holley:2023,Robinson:2023}. In the US, multiple recent announcements from the National Science Foundation have explicitly emphasized equity and the inclusion of marginalized groups in access to quantum education programs \cite{Marrongelle:2020,NSF:2020}. Meanwhile, the Quantum Information Science and Technology Workforce Development National Strategic Plan states,

\begin{iquote}
    ``As the need for [QIS] expertise grows, we must ensure that considerations of diversity, equity, and inclusion play a key role in all developments.'' \cite{SCQIS:2022}
\end{iquote} The report establishes a bold vision for QIS education in the United States:
\begin{iquote}
    ``\textbf{Vision:} The United States should develop a diverse, inclusive, and sustainable workforce that possesses the broad range of skills needed by industry, academia, and the U.S. Government, while being able to scale and adapt as the [QIS] landscape evolves.'' \cite{SCQIS:2022}
\end{iquote} The report also articulates a series of associated actions, most notably:
\begin{iquote}
    ``\textbf{Action 4:} Make careers in [QIS] and related fields more accessible and equitable.'' \cite{SCQIS:2022}
\end{iquote}

From these statements, it is clear that quantum policymakers -- and indeed large parts of the quantum community as a whole -- claim to deeply prioritize equity and access of quantum education programs. Of course, terms such as \textit{equity} and \textit{access} are seldom explicitly defined by the community, and each individual and agency may well come to their own understanding of what these terms mean. For the purpose of our analysis and discussion, we adopt the following definitions of \textit{equality}, \textit{equity}, and \textit{access}:

\begin{itemize}
    \item \textbf{Equality:} Quantum education programs should be made equally available to Americans of all communities and social backgrounds. Quantum education should not be a luxury restricted only to the wealthy or historically advantaged.
    \item \textbf{Equity:} In the presence of ongoing injustice throughout the educational pipeline and broader society, it can in fact be argued that marginalized students may in fact deserve greater access to resources simply to facilitate comparable outcomes \cite{Espinoza:2007}. A focus on outcomes, rather than simply equal treatment in an otherwise unequal system, thus differentiates the goals of equality and equity \cite{Menendian:2023}. 
    \item \textbf{Access:} Any American who desires to study QIS and possesses the willingness to learn the necessary skills should have the opportunity to do so. Quantum education should be made a possibility for individuals from as many walks of life as possible, including those already embedded in specific educational, cultural, or geographic communities.
\end{itemize}

A natural corollary to these goals is that \textbf{students should be equally able to access the benefits of quantum education wherever they go to school}. Perhaps fully realizing this vision is impractical. However, if the quantum community wishes to take the commitment to equity seriously, we must at least consider the implications of this vision and what they mean for educational policy.

In this paper, we seek to determine what underlying factors make an institution more or less likely to have QIS education programs today, while the push for QIS education remains in its formative years. 
In other words, we wish to answer the research question, ``Where are US postsecondary QIS educational programs being offered, and based on this, is access to these educational programs equitable for all students?''

We emphasize that our quantitative analysis can only identify equality, not equity, gaps in access to QIS education. Indeed, it is unclear how any such purely quantitative study \textit{could} directly attempt to measure equity. However, in our analysis we consistently observe that institutions serving generally less privileged (e.g.\ low-income or rural) student bodies also offer fewer QIS educational programs. The consistent sign of these numerical disparities makes the equality gaps we discuss not just an equality, but an equity, concern, as we discuss in further length in Sec.~\ref{sec:conclusions}.

Our methods also cannot determine the underlying cause of any disparity we observe, nor can they indicate (at least until follow-up studies can be conducted) whether access to QIS education is becoming more or less equitable over time. What our work does aim to do is to quantify whether or not certain classes of institutions -- and the students that attend them -- are more likely to offer quantum education programs than others. When allocating resources, educators and policymakers alike may wish to ask: Does the current distribution of QIS education programs align with our goal of equitable access to quantum careers -- and if not, how can we change it?

\section{Methodology}

To quantitatively investigate the distribution of QIS coursework and degree programs at US institutions, we compiled a comprehensive database of QIS courses and degrees offered across 456 US colleges and universities as of August-September 2022.

\subsection{Selection of institutions}
\label{sec:inst-selection}

Previous work has shown that QIS coursework tends to be primarily concentrated in physics, computer science (CS), and electrical and/or computer engineering (ECE) departments \cite{Meyer:2022PhysRev}. As such, we expect that QIS coursework will generally only be offered at institutions that have sizable programs in one or more of these disciplines. Since one purpose of our catalog search was to identify instructor emails for subsequent distribution of a faculty survey \cite{Meyer:2023IEEE}, it was not practical to analyze a truly random sample of US colleges and universities. Instead, similar to Ref.~\cite{Cervantes:2021}, our catalog search narrowly focuses on those institutions deemed sufficiently likely to have a QIS course to merit further examination. We anticipate that this methodological choice will, if anything, tend to underestimate the size of disparities across classes of institutions: inclusion in our catalog search is already restricted only to those institutions with sizable STEM departments, and the distribution of STEM departments themselves is potentially influenced by the factors we consider in our analysis. Appendix~\ref{app:representative} explores these possible effects in greater depth. 

Our catalog search includes all institutions searched in Ref.~\cite{Cervantes:2021}, as well as those meeting one or more of the following criteria in 2021 for any of the fields of physics, CS, or ECE\footnote{We calculate the number of degrees for each discipline by aggregating the number of degrees awarded under the following 4-digit CIP (Classification of Instructional Programs) codes:

\textbf{Physics:} 40.08 (physical sciences - physics), 14.12 (engineering - engineering physics)

\textbf{CS:} 11.07 (computer and information sciences and support services - computer science)

\textbf{ECE:} 14.09 (engineering - computer engineering), 14.10 (engineering - electrical, electronics, and communications engineering)
} per the publicly-available National Center for Education Statistics' IPEDS database \cite{IPEDS}\footnote{These same criteria were also used to identify recipients for a QIS faculty survey reported in Ref.~\cite{Meyer:2023IEEE}}:

\begin{itemize}
    \item Top 100 bachelor's degree-granting program in field, and/or awarded 30 or more bachelor's degrees in field
    \item Top 50 master's degree-granting program in field, and/or awarded 15 or more master's degrees in field
    \item Top 50 Ph.D.-granting program in field, and/or awarded 10 or more research-based Ph.D.'s in field
\end{itemize}

These classifications encompassed a total of 475 institutions. For reasons of data availability and small numbers of institutions, we opted to exclude private for-profit institutions and military academies from analysis for the purposes of this paper. For the same reasons, we restricted our dataset to universities covered by Title IV\footnote{\label{fn:title-iv}A small number of institutions opt out of federal data collection under Title IV, generally citing religious regions \cite{Caputo:2016}. Non-Title IV institutions cannot accept federal financial aid.} and located in the 50 US states and the District of Columbia (DC). These restrictions slightly reduced the number of institutions analyzed for the purpose of this study to $N=456$.

\subsection{Identification and classification of QIS coursework}

Data on available courses varied somewhat from institution to institution. To ensure the thoroughness of our search, we cross-referenced the following publicly-available databases:

\begin{itemize}
    \item Digitized course catalogs (most recent available catalog as of August-September 2022, either 2021-2022 or 2022-2023 academic year)
    \item Where available, current or archived course schedules dating back to fall term 2019 for standard terms during the academic year, excluding summer (including published anticipated course schedules through spring term 2023 if available)
    \item Where available, current and archived course titles on Coursicle \cite{Coursicle} dating back to fall 2019 or the start of data availability
\end{itemize}

On Coursicle and in some institutions' course catalogs and schedules, course listings were published in separate documents for each department. In such cases, only the catalogs for physics, engineering physics, mathematics, computer science, and electrical and/or computer engineering (or general engineering if engineering subdisciplines were not classified separately) were searched, alongside any department with ``quantum'' in the name.

To identify possible QIS courses, we searched course titles -- and where available, course descriptions -- for the word ``quantum''\footnote{Our procedure may miss cases in which the word ``quantum'' is abbreviated, particularly if the abbreviation is nonstandard. We minimized this concern by cross-referencing across multiple databases and searching course descriptions (not just titles) wherever possible. In a few cases where only heavily-abbreviated course titles were available, we resorted to manually searching all course titles in the disciplines of interest that contained the letter ``Q.''} and classified each course according to the following coding schema. Courses identified as QIS include those meeting any of the following descriptions, provided that core QIS topics are judged to comprise 50\% or more of the content:\footnote{For instance, a laboratory experiment on a core QIS topic, such as Bell's inequalities, would count toward the 50\% threshold, but a lab simply intended to build general optics skills that an interested student could hypothetically one day find useful for QIS applications would not count. Similarly, a course on hardware and materials would only count if 50\% or more of the course was about device fabrication specifically for QIS applications, whereas a course simply on manufacturing photonics devices that could just as easily be used for a variety of non-QIS applications would most likely not qualify.}

\begin{itemize}
    \item \textbf{QIS theory}: A theory course covering one or more aspects of quantum technology -- quantum computing (including quantum simulation), quantum communications and networking (including post-quantum classical cryptography), or quantum sensing and metrology. May be restricted to specific applications or an advanced subtopic. May include one or more related topics, such as quantum optics, as long as QIS represents 50\% or more of the covered content.
    \item \textbf{Lab/practicum - QIS}: A laboratory or technical practicum focused primarily on QIS applications
    \item \textbf{1-2 unit seminar - QIS}: A short seminar primarily covering aspects of QIS theory or practice, presumably without a final exam or extensive assignments
    \item \textbf{Math methods for QIS}: A mathematics class specifically advertised as preparing students for future work in QIS or quantum computing
    \item \textbf{Quantum mechanics for QIS}: An applied quantum mechanics course specifically advertised as preparing students for future work in QIS or quantum computing, as distinguished from a general quantum mechanics course offered to physicists or electrical engineers
    \item \textbf{Hardware/materials for QIS}: A course focused on the development of hardware for QIS applications, including  fabrication and characterization of materials.
    \item \textbf{Non-STEM - QIS}: A non-STEM course focusing on the applications or societal impacts of quantum technologies as at least 50\% of the course. (Only 2 such courses were identified.)
\end{itemize}

Survey courses in physics, computer science, engineering, or other disciplines that included a brief unit on QIS were not included unless QIS was judged to make up 50\% or more of the core content of the course. Likewise, traditional quantum mechanics or modern physics courses with 
a unit on one or more QIS applications were excluded unless the 50\% threshold was deemed met. Independent study, directed reading, and independent research courses were categorically excluded even if they otherwise met one of the above definitions. Multiple listings of the same course across departments or undergraduate/graduate levels were collapsed into a single entry.

To avoid counting obsolete courses remaining as relics in college catalogs, courses were classified as ``active'' and counted only if either (a) the course was newly added to the institution's catalog in the 2019-2020 academic year or later or (b) we could confirm that the course was offered at least once since fall 2019. Recent offering of a course was typically confirmed through archived course schedules or Coursicle, although alternative evidence that the course was offered -- such as archived syllabi or a statement on a faculty member's CV or departmental website -- was sometimes relied upon when schedules were unavailable.

Special topics (ST) courses offered on an experimental basis posed a particular challenge for our search. Data on such courses was available from some institutions but not others, and tended to be most readily available from large R1 institutions. Confounding the problem, for some institutions special topics course information was available from certain semesters or departments but not others. We catalogued these courses where available but removed them from our dataset for regression analysis because they could not be reliably counted.

To test interrater reliability, a random sample of 40 active, non-special-topics courses containing a QIS keyword (such as ``quantum computing'' or ``qubits'') in the course description were assigned to a second rater, who was asked to rate each course as QIS or not QIS per the initial version of the above codebook. Initial percentage agreement was an unacceptably low 80\%. After one round of minor modifications to the codebook (including the addition of the text reproduced here in footnote 5), an acceptable interrater agreement of 95\% (Cohen's $\kappa = 0.90$) was achieved.

\subsection{Identification and classification of QIS degrees}

We likewise conducted a search for QIS degree and certificate programs at those institutions that were deemed to have sufficient coursework to potentially support a degree or certificate program: either (a) at least two active, non-ST QIS courses or (b) one active, non-ST QIS course and 2 or more active special topics QIS courses. We searched the overall lists of degrees and certificates for each institution, as well as concentrations in physics, CS, ECE, and any other department found to have QIS courses identified in our catalog search. 
We only included programs that were currently active or had received final approval as of our search in August-September 2022. We also excluded noncredit online programs with noncompetitive admissions. We do include named QIS-specific concentrations housed within a broader degree program, provided that these concentrations require students to take additional or substitute QIS-specific coursework not required for the principal degree track (as opposed to simply conducting thesis research in QIS).

We identified a total of 45 degree and certificate programs across 35 of the 456 institutions surveyed. For reference, a complete list of the degree and certificate programs is provided in Appendix~\ref{app:list}.

\subsection{Poisson and negative binomial regression}
\label{sec:regression-methodology}

We wish to model the number of QIS courses (or number of degree and certificate programs) as a function of institutional factors. Our dependent variable (number of courses) is therefore a discrete, non-binary count variable. While such count variables may be relatively rare in educational data, approaches from the biomedical and social sciences are well-suited to this type of analysis. 

The simplest discrete count model is Poisson regression \cite{Coxe:2009}, which models the dependent variable as a Poissonian random variable whose mean is a function of the independent variables. In physics, Poisson processes are commonly encountered when modeling counts per unit of time that occur due to a stochastic process, such as radioactive decays. For the purposes of this analysis, the ``unit'' we are considering is one institution of higher learning and the counts are number of courses (or degrees).

Standard Poisson regression models the logarithm of the mean number of courses or degrees, $\mu$, by a simple linear combination of the independent variables $\vec\theta$ (with possible interaction terms).
A key assumption of the Poisson model is that the conditional mean $\mu(\vec\theta)$ equals the conditional variance $\sigma^2(\vec\theta)$ for any set of parameters $\vec\theta$. For real social science datasets, the condition of equal mean and variance may not be met; instead, real-world data often exhibits significant overdispersion [$\mu(\vec\theta) > \sigma^2(\vec\theta)$] due to the inevitable presence of unobserved, uncontrolled causal variables. Simple Poisson regression on overdispersed data tends to produce inflated $p$ values and thus alternate methods are preferred where a $t$ test of Poisson regression residuals reveals significant overdispersion.

Various alternatives to simple Poisson regression have been developed to correct for specific causes of overdispersion (e.g.\ excess zeros) \cite{Coxe:2009}. The simplest and most commonly-used such technique -- which we adopt -- is Negative Binomial (NB) regression \cite{VerHoef:2007, Yang:2015} which models the variance as a generalized function of the mean:

\begin{equation}
    \sigma^2(\vec\theta) = \mu(\vec\theta) + k_q \mu^q (\vec\theta)
    \label{eq:negbin-dispersion}
\end{equation}
Here, $q$ is a power equal to either 1 (Type I NB) or 2 (Type II NB) and $k_q$ is an undetermined overdispersion parameter which is fit as part of the regression model.\footnote{The variable we call $k_q$ is typically referred to as $\alpha$ in the literature. We adopt the notation $k_q$ to avoid confusion with significance level $\alpha$ and emphasize that $k_1$ and $k_2$ values in respective models are not directly comparable.} See e.g.\ Refs.~\cite{Hadi:1995,Liu:2005,Piza:2012} for various applications of negative binomial regression in the literature.

For each of the models discussed in this analysis, we first performed a simple Poisson regression and performed a $t$ test on the residuals to determine whether statistically significant overdispersion was present; we then used Negative Binomial regression in cases of overdispersion. If negative binomial regression was warranted, we empirically selected $q$ for our analysis based on whichever choice of model produced better fit as determined by the Bayesian Information Criterion.\footnote{Since our goal of statistical modeling here is purely descriptive rather than causal, prioritizing model fit for the same number of parameters is warranted.} See Appendix~\ref{app:flowchart} for a more detailed of the regression process described in this section.

\begin{table*}[t]
    \centering
    \begin{tabular}{ l  l  l }
        \hline \hline
         \thead{Independent variable} &  \thead{Indicator vars} & \thead{Definition}\\
        \hline \\

        Carnegie classification & \makecell[l]{Is\_R1\\Is\_R2} & \makecell[l]{$=1$ if an R1 institution, 0 otherwise \\ $=1$ if an R2 institution, 0 otherwise} \vspace{4pt}\\ 

        Funding & Is\_Public & $=1$ if public, 0 if private not-for-profit \vspace{4pt}\\

        Religious Affiliation & Is\_Religious & $=1$ if current religious affiliation, 0 otherwise \vspace{4pt}\\

        URM serving & Is\_URM\_Serving & $=1$ if MSI (not solely AANAPISI), 0 otherwise \vspace{4pt}\\

        Low-income serving & Pell\_Grant & $=$ \% undergrads receiving Pell Grant, $z$-transformed \vspace{4pt}\\

        Urbanization & Urban\_Index & $=$ FiveThirtyEight urbanization index, $z$-transformed \\
        \hline \hline
        
    \end{tabular}
    \caption{Independent variables included in regression analysis, along with associated indicator variables for categorical data.}
    \label{tab:indicators}
\end{table*}

\subsection{Independent variables we consider}
\label{sec:vars}

We consider a variety of institutional factors as potential independent variables that may correlate with the number of courses or degrees offered by an institution. The choice of independent variables for regression analysis is based in large part on the observations of Ref.~\cite{Cervantes:2021} which found potential disparities along the lines of Carnegie classification, funding, and status as underrepresented minority (URM) serving institutions. We also include variables to test for the effects of poverty (percentage of students receiving Pell grants) and access for students in rural areas. In this section, we define each of the independent variable we consider. The following section (Sec.~\ref{sec:vars-explanation}) motivates the inclusion of each of these variables in terms of their significance with respect to possible equity concerns.

\begin{outline}
    \1 \textbf{Carnegie classification}: The college or university's basic classification according to the 2021 Carnegie Classification of Institutions of Higher Learning \cite{Carnegie}. We bin the institution's basic classification into 3 categories:
    \2 R1 (Doctoral universities, very high research activity)
    \2 R2 (Doctoral universities, high research activity)
    \2 Non-research (Doctoral universities not in either of the above classifications, as well as master's and baccalaureate institutions)

\vspace{10pt}

    \1 \textbf{Funding}: Public or private-not-for-profit.

    \1 \textbf{Religious affiliation}: Presence or absence of a current religious affiliation. (In practice, in each case we consider, this variable drops out of our statistical models once other variables are controlled for.) 

    \1 \textbf{MSI status}: Whether the institution is a minority-serving institution as defined by NASA's 2020 List of Minority Serving Institutions \cite{NASA:2020}. 
    We exclude institutions that are classified solely as AANAPISI (Asian American and Native American Pacific Islander-serving institutions\footnote{\label{fn:aanapisi}``Native American Pacific Islander'' is defined in the relevant statute as ``any descendant of the aboriginal people of any island in the Pacific Ocean that is a
territory or possession of the United States'' \cite{AANAPISI-def}. (Contrary to the ordinary usage of the term ``Native American,'' the AANAPISI classification is \textit{not} intended to include descendents of peoples indigenous to mainland North or South America.)}) from this categorization given that Asian Americans tend to be well-represented in STEM fields and are presumed to be the principal minority bloc at AANAPISIs that do not meet the criteria for any additional MSI code.\footnote{A separate code exists specifically for institutions serving at least 10\% Native Hawaiian students \cite{NHSI-def}. We are operating on the assumption that an AANAPISI institution whose primary minority bloc is Pacific Islander and which is located in the 50 US states or D.C. will most likely also be classified as Native Hawaiian-serving.}

    \1 \textbf{Percentage of undergraduate students receiving a Pell grant}: As a proxy for the financial resources of the student body, we consider the percentage of undergraduate students at the institution receiving a need-based Pell grant in 2020-2021. We rescale according to a standard $z$ transformation, yielding a mean of 0 and standard deviation of 1 across all institutions in the dataset.

       \1 \textbf{State urbanization index}: As a proxy for regional urbanization, we adopt the FiveThirtyEight urbanization index \cite{Rakich:2020}, which was developed for political science research as an indicator of state-level urbanization. The index represents the natural logarithm of the average number of people living within 5 miles (8.0 km) of any resident of the state. We rescale using a $z$ transformation to produce a mean of 0 and standard deviation of 1 across the 50 states plus DC.

\end{outline}

\noindent Independent variables included in the regression analysis and associated categorical indicator variables are summarized in Table~\ref{tab:indicators}.
Per standard practice for Poisson and negative binomial regression, we model the logarithm of the mean number of courses or degrees $\mu$ by a simple linear model of the indicator variables $\vec\theta$. We also include interaction terms up to second order, involving at least one continuous variable:
\begin{equation}
    \ln (\mu(\vec\theta)) = a_0 + \sum_{i}\beta_i \theta_i + \sum_i\sum_{\substack{j>i \\ \theta_j \text{cont.}}}\gamma_{ij}\theta_i \theta_j
\end{equation}
where $a_0$, $\{\beta_i\}$ and $\{\gamma_{ij}\}$ are the undetermined regression coefficients. We iteratively remove terms from the regression model in order from greatest to least $p$ value until the remaining coefficient(s) of highest order in each $\theta_i$ are significant at the $\alpha=.05$ level with Holm-Bonferroni correction for multiple simultaneous statistical tests.\footnote{This choice of cutoff in the iterative deletion process also optimizes the Bayesian information criterion -- a common metric for regression model goodness of fit that penalizes overfitting -- for our models of both courses and degrees, providing an additional sanity test that our model is appropriately fit.} Since fully 80\% (28/35) of the institutions with at least one QIS degree or certificate program were R1 institutions, we also excluded cross-terms involving the variable is\_R1 in our analysis of degree and certificate programs to avoid spurious correlations between independent variables. This procedure is summarized in Appendix~\ref{app:flowchart}.

\begin{figure*}[htb]
    \centering
    \subfigure[][\ Percentage of enrolled undergraduates from top 1\% of family income distribution ($\ge$\$630k/yr).]{
        \includegraphics[scale=1]{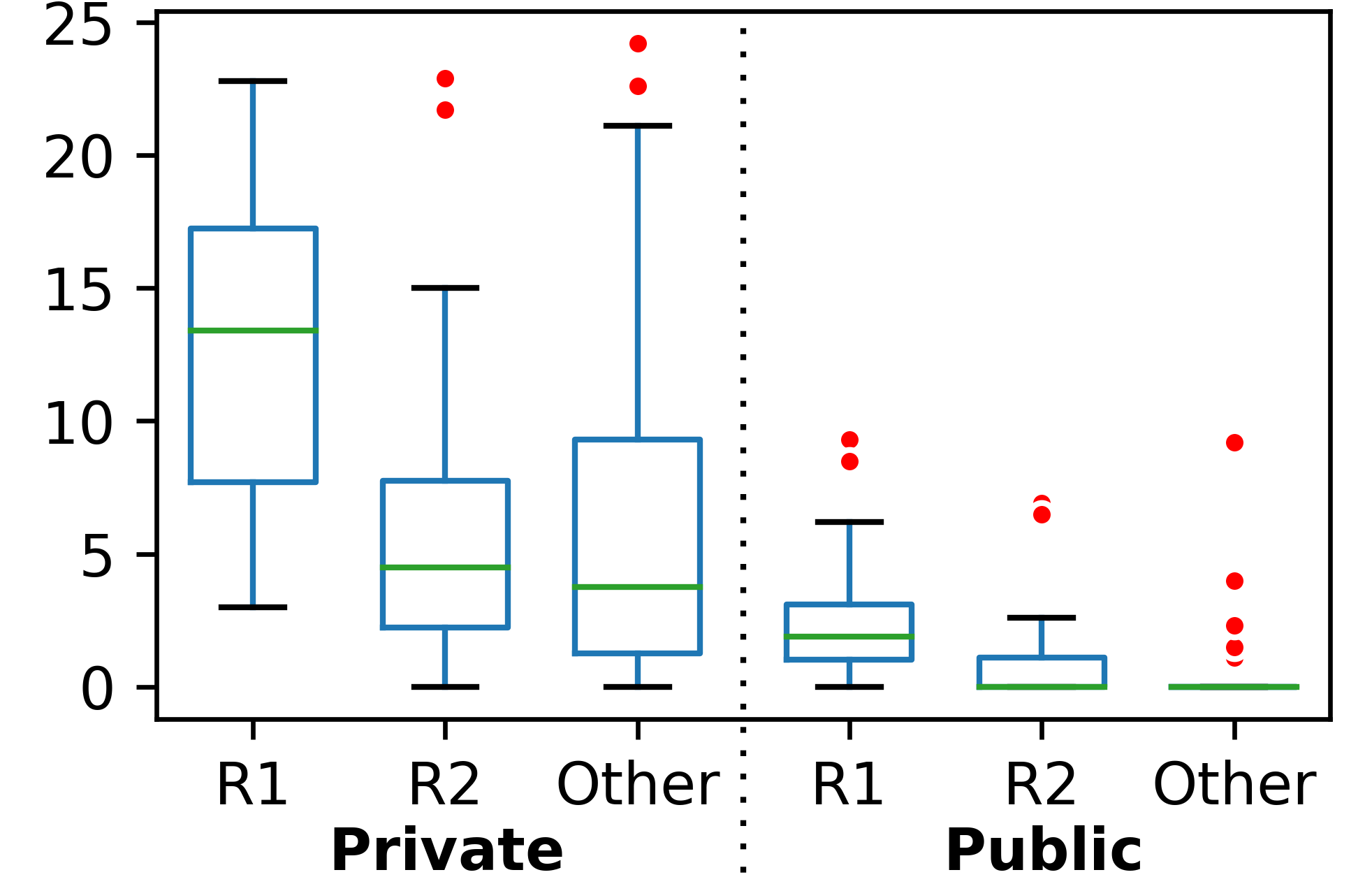}
}
    \subfigure[][\ Percentage of enrolled undergraduates from bottom 60\% of family income distribution (<\$65k/yr).]{
        \includegraphics[scale=1]{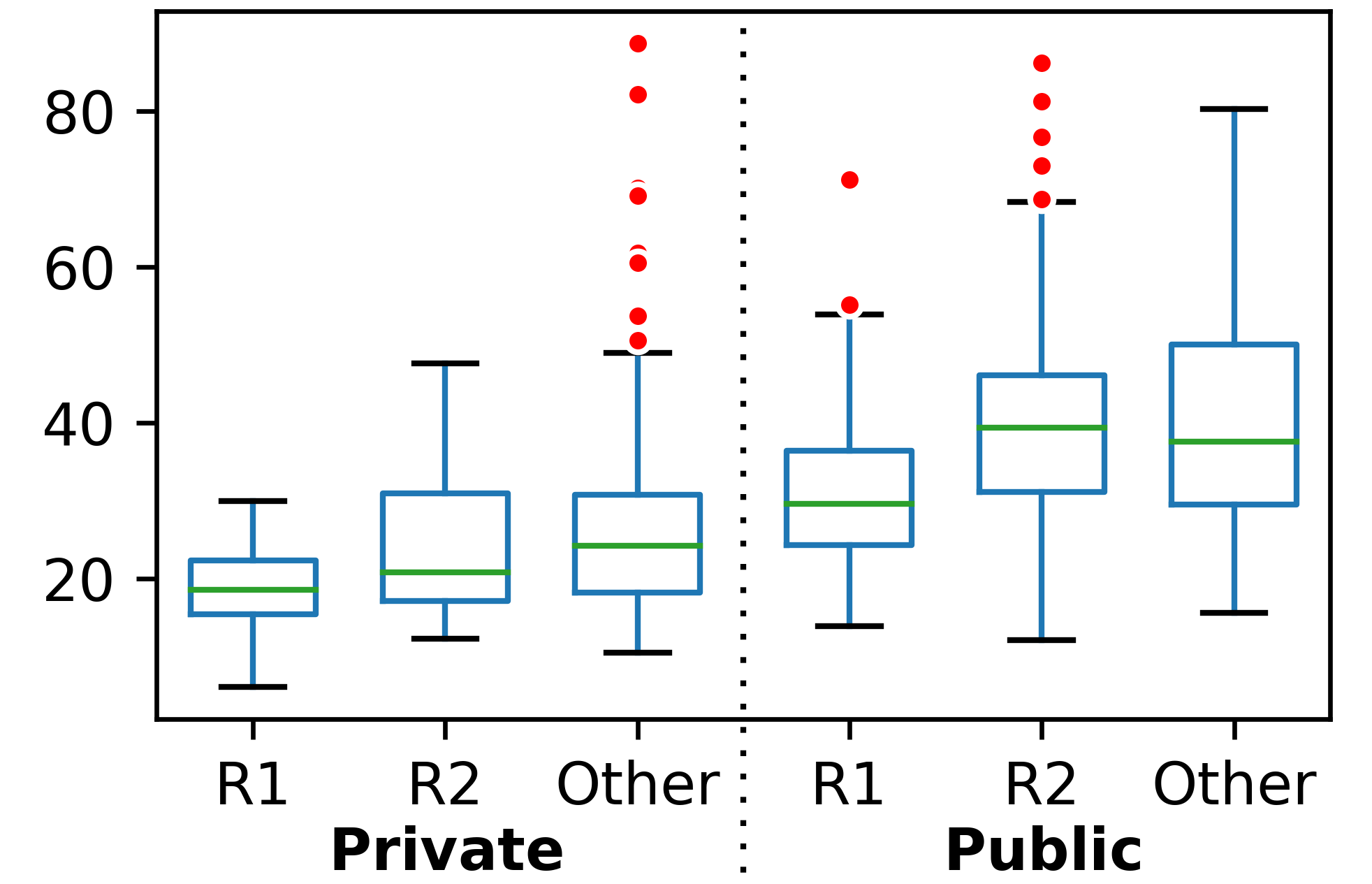}
        }
    \caption{Percentage of enrolled undergraduate student body in (a) top 1\% vs.\ (b) bottom 60\% of family income distribution among institutions investigated in our study, by Carnegie classification and funding (public vs.\ private). Numbers taken from a 2017 study of anonymized tax data \cite{Chetty:2017, NYT:2017} and are for domestic students in the age 18-22 bracket (typically first-time undergraduates). Data is for 1992 birth cohort (approximately class of 2013). Data available for 387 (85\%) of the $N=456$ institutions investigated as part of this study. Red dots represent possible outlier institutions (percentage of students from respective income bracket differs from the median by more than 1.5 times the inter-quartile range, represented by the height of the blue rectangle). 
    }
    \label{fig:income}
\end{figure*}

\subsection{Why these independent variables?}
\label{sec:vars-explanation}

\begin{figure*}[htb]
    \centering
    \subfigure[][\ Annual net costs of attendance for low- to moderate-income undergraduates at median institution in classification (USD/yr).]{
        \includegraphics[scale=.95]{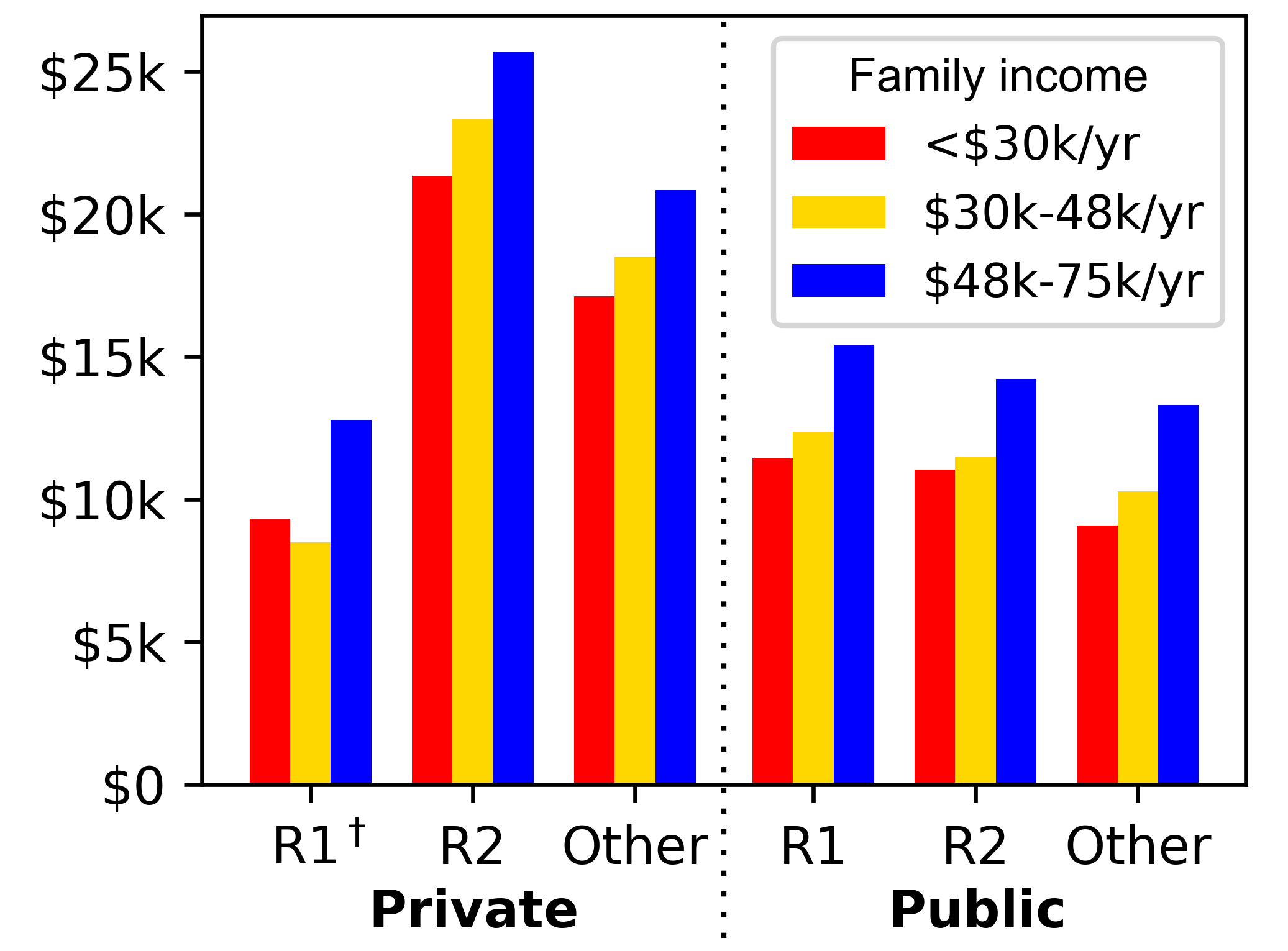} 
        \label{fig:costs-attendance}}
    \subfigure[][\ Admissions selectivity (percent of undergraduate applicants admitted, 2021 cycle*) by funding and Carnegie classification.]{
            \includegraphics[scale=.95]{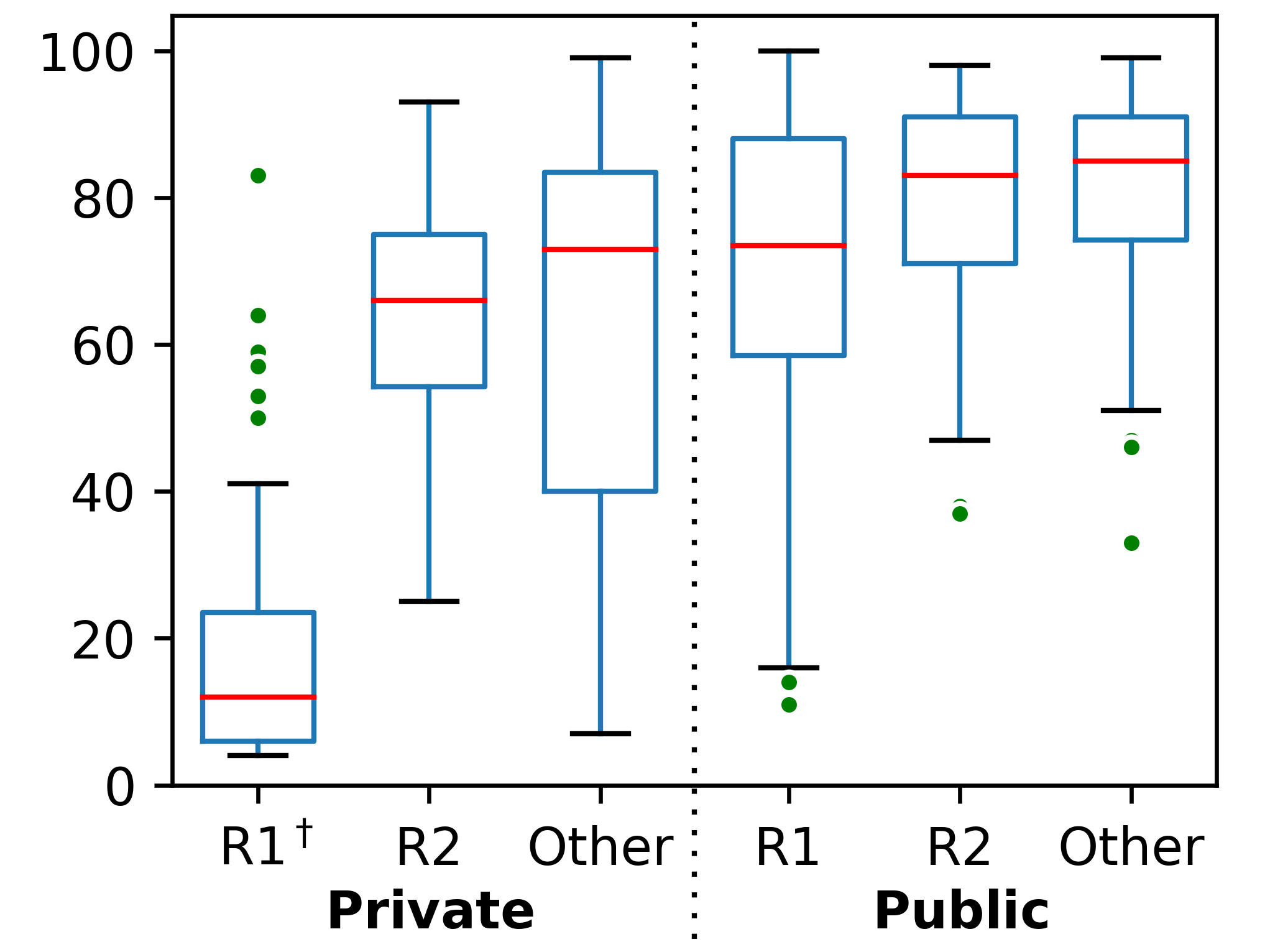}
            \label{fig:selectivity}}

    \caption{Low- to moderate-income students attending private universities pay higher annual out of pocket costs. While private R1 institutions at first appear an exception -- indeed, many are known for giving generous financial aid to the few low- to moderate-income students they do admit \cite{Berke:2016} -- these institutions' high selectivity makes them unreachable for all but the most academically-accomplished low- and moderate-income students. \footnotesize\newline \\ $^\ast$Percent admitted is of the applicant pool, not the entire US population; the application process itself -- typically involving fees and extensive essays -- already limits applicant pools at selective institutions only to students believing they have a reasonable probability of admission to begin with. \newline $^\dagger$As represented by the green dots in (b), a few outlier private R1 institutions with relatively high acceptance rates do exist; however, such institutions generally have costs of attendance similar to those of private non-R1s. Notably, a student with a family income below \$30k/yr will pay a median of \$17500/yr at private R1s with an admit rate above 20\% vs.~\$5600/yr at private R1s with an admit rate below 20\%. }
    \label{fig:costs-attendance-total}
\end{figure*}

\begin{outline}
    \1 \textbf{Carnegie classification and funding:} In general, as shown in Fig.~\ref{fig:income}, R1 institutions enroll a markedly wealthier student body, particularly at the undergraduate level, than non-research institutions, with R2 institutions occupying a middle tier. Likewise, private institutions enroll wealthier student bodies than public institutions. The gap in enrollment of low-to-moderate income students may be explained by differences in cost of attendance -- particularly for low-income students -- as shown disaggregated by family income in Fig.~\ref{fig:costs-attendance}. Thus, if in practice students at private research institutions have access to a greater number of QIS educational opportunities than their peers at other classes of institutions, then current QIS education programs are disproportionately benefiting already privileged students.

    \1 \textbf{Religious affiliation:} 
    Public institutions are required to be secular under US law. Considering religious affiliation in our model helps establish that any disparity along the axis of funding -- a potential equity concern -- is not actually a proxy for religious affiliation.

    \1 \textbf{MSI status:} Ref.~\cite{Cervantes:2021} identified a potential disparity in access to QIS coursework at minority-serving vs.\ predominantly-white institutions, a finding of particular concern for racial equality (and, given the direction of the disparity, racial equity).

    \1 \textbf{Pell grant:} Need-based Pell grants are typically reserved for undergraduate students from families in approximately the lower 40\% of the income distribution. The percentage of students receiving a Pell grant is thus a proxy for the financial resources of the student body.

    \1 \textbf{Urbanization index:} 
     Rural students in the United States face a number of barriers to higher education that cannot always be framed strictly in terms of race or class. Rural students may lack exposure to and support for STEM careers at home and in high school \cite{Martinez:2003,Boynton:2012}, and rural schools often struggle to recruit quality STEM teachers \cite{Ivey:2016} or offer advanced coursework \cite{Beckwith:2020}. Rural students have reported struggling to fit in culturally on large campuses and in urban areas \cite{Beckwith:2020}. Marginalized rural students in STEM, such as Hispanic students \cite{Lopez:2022} and women pursuing engineering \cite{Vaziri:2020}, face unique resource and educational challenges at the intersection of rural status and other identities. Students from rural America are pursuing higher education at declining rates compared to urban and suburban America \cite{Marcus:2017}, and many that do attend college remain close to home even if degree offerings are limited \cite{Hillman:2019}. While particularly motivated students might, of course, choose to attend school out of state in order to access quantum programs, doing so risks exacerbating ``brain drain'' from rural to urban areas
       \cite{Boynton:2012}. Ref.~\cite{Cervantes:2021} found that QIS coursework seems to be disproportionately located at institutions near the coasts and other highly urbanized states such as Texas, a concern for rural-urban equality.
       \vspace{10pt}
       
       We adopt a state-level indicator for urbanization because institutions of higher learning tend to draw from their broader region in terms of enrollment and collaborations. For instance, the University of Illinois at Urbana-Champaign presumably benefits from regional proximity to Chicago even though it is located in a small-town setting, and likewise students from Chicago attending the school can receive in-state tuition.\footnote{We note that in the present U.S.\ political climate, urbanization is also a close proxy for state political orientation; Rakich \textit{et al.} found a correlation of 0.69 between FiveThirtyEight urbanization index and presidential vote in 2016 \cite{Rakich:2020}. However, while our analysis cannot fully disentangle urbanization from partisanship, we argue that urban-rural disparities themselves present an equity concern even if the effect is partially mediated by political control. Moreover, the bipartisan support of the National Quantum Initiative Act of 2018 \cite{NQIA:2018} does not suggest quantum technologies are a site of particular political polarization.}
\end{outline}

\begin{figure*}[htbp]
    \centering
    \includegraphics[scale = 0.75]{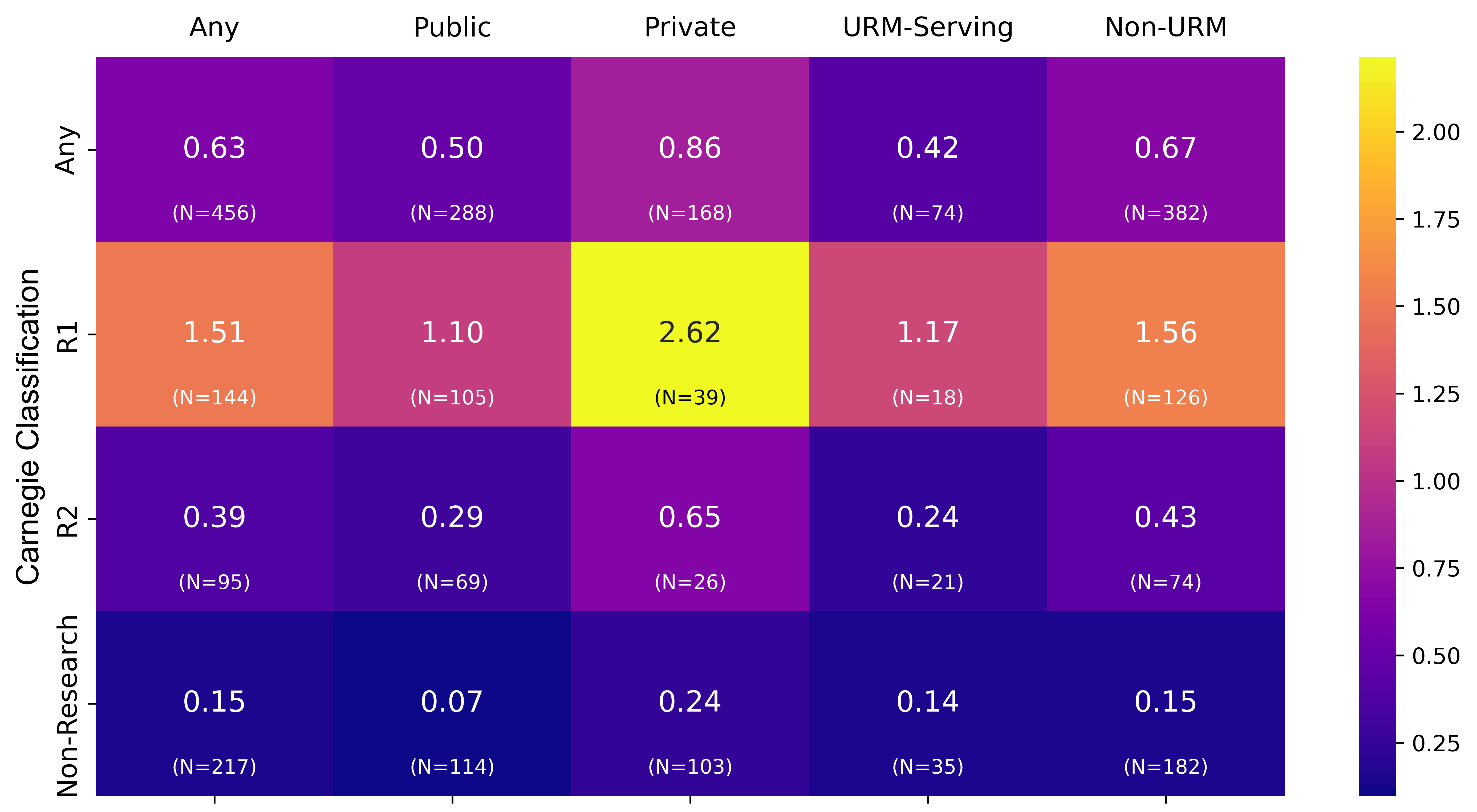}
    \caption{Mean number of courses offered at institution, by Carnegie classification, funding (private/public), and status as an URM-serving institution. The mean number of courses per institution across the entire dataset is $\mu = 0.63$ (median 0), with a standard deviation of $\sigma = 1.39$.}
    \label{fig:heatmap}
\end{figure*}

\section{Distribution of QIS coursework}

\subsection{Qualitative analysis}

Figure~\ref{fig:heatmap} shows the mean number of courses offered by Carnegie classification. Immediately, we see that QIS courses are disproportionately likely to be found at R1, private, secular institutions. URM-serving institutions are also somewhat less likely to offer QIS courses than non-URM-serving institutions, though the effect is less obvious. It is worth noting that despite the recent surge in QIS coursework, the median number of QIS courses identified at an institution is 0 and the mean is 0.63, indicating that most institutions do not yet have any formalized QIS courses.

\subsection{Regression results}
\label{sec:courses-regression}

As discussed in Sec.~\ref{sec:regression-methodology}, we use a NB regression model to quantify the effect of each independent variable on the mean number of QIS courses offered at an institution. Results from the regression analysis are summarized in Table~\ref{tab:results-regression}. 
Our regression results reveal a number of important conclusions:
\begin{table*}[htbp]
    \centering
    \begin{tabular}{c r c c c c c }
    \hline \hline
         \thead{Term} &  \thead{Coefficient} & \thead{$p$} & \thead{Significance} & \thead{95\% CI} & \thead{\makecell{Rate ratio$^\dagger$}} & \thead{95\% CI}\\
        \hline
         Intercept ($a_0$) & $-1.75 \pm .27$ & $p < .001$ &  \s\s\s & [-2.27, -1.23] & -- & -- \\ 
         Dispersion ($k_1$) & $1.11 \pm .26$ & $p < .001$ & \s\s\s  & [0.60, 1.61] & -- & --
         \\ \hline
         Is\_R1 & $\ 2.18 \pm .25$ & $p < .001$ & \s\s\s & [1.70, 2.66] & 8.8 & [5.5, 14.3]  \\
         Is\_R2 & $1.01 \pm .30$ & $p < .001$ & \s\s & [0.41, 1.60] & 2.7 & [1.5, 5.0]  \\
         Is\_Public & $-0.83 \pm .25$ & $p<.001$ & \s\s & [-1.33, -0.34] & $2.3^{-1}$ & [$1.4^{-1}$, $3.8^{-1}$] \\
         Pell\_Grant & $-0.39 \pm 0.14$ & $p=.006$ & \s & [-0.65, -0.11] & $1.5^{-1}$ & [$1.1^{-1}$, $1.9^{-1}$]  \\
         Urban\_Index & $0.13 \pm .14$ & $p=.35$ & -- & [-0.14, 0.40] & -- & -- \\ 
         Public $|$ Urban & $0.67 \pm .21$ & $p = .001$ & \s\s & [$0.26$, $1.08$] & $2.0$ & [$1.3$, $2.9$] \\ \hline \hline
         
    \end{tabular}
    \caption{Final regression results for modeling number of active QIS courses at each institution (pseudo-$r^2$ = .17). Vertical bar denotes interaction terms (i.e.\ Public | Urban represents the interaction between Is\_Public and Urban\_Index). Type I NB regression ($q=1$ in Eq.~\ref{eq:negbin-dispersion}). Significance calculated using Holm-Bonferroni adjustment for multiple statistical tests: \\ \s \ $p_{adj} < .05$ \ \
    \s\s \ $p_{adj} < .01$ \ \
    \s\s\s \ $p_{adj} < .001$\\
    \footnotesize{\\ $^\dagger$ Rate ratio is an effect size measure representing the multiplicative effect on the mean number of courses offered at an institution. For interaction terms between a continuous and an indicator variable, rate ratio is over a change by 1 standard deviation in the value of the continuous variable. In the absence of consensus guidance in the literature, we follow Ref.~\cite{Olivier:2017} in interpreting a rate ratio of 1.2 as a small effect, 1.9 as a medium effect, and 3.0 as a large effect.}
}
    \label{tab:results-regression}

\end{table*}

\begin{itemize}
    \item QIS courses and degrees are disproportionately found at large research universities. This effect is very large: all other factors being equal, a student at a typical non-research institution has access to nearly 9 times fewer QIS courses as a student at a typical R1 university, and 3 times fewer than a student at a typical R2. As seen in Fig.~\ref{fig:heatmap}, the mean R1 institution has 1.5 QIS courses while the mean non-research institution has only 0.15; the disparity seen in the regression analysis simply confirms that QIS coursework is very rare at non-research institutions. These observations echo Refs.~\cite{Perron:2021,Cervantes:2021}.

    \item Likewise, large disparity exists between private and public institutions for both total courses and degree/certificate programs. No statistically significant difference was found on the axis of religious affiliation among private institutions, once other factors are controlled for.

    \item Institutions with a high percentage of undergraduates receiving Pell grants offer significantly fewer QIS courses, indicating that low-income students may have fewer opportunities to access QIS coursework. This effect is \textit{in addition to} any gaps in access along the axis of income due to stratification of enrollment by Carnegie classification and funding. A similar gap is observed in access to QIS degrees at public institutions.

    \item We find a moderate positive association between state-level urbanization and number of QIS courses and degrees offered. For both overall courses and degrees, this gap is specific to publicly-funded institutions, where rural students can access benefits such as in-state tuition.

    \item We find no statistically significant correlation between status as a URM-serving institution and number of QIS courses offered, once other factors are controlled for.
    
\end{itemize}

The results are qualitatively similar when looking specifically at the number of QIS courses available to \textit{undergraduates} (Table~\ref{tab:results-regression-UG}), for whom the distribution of courses is perhaps even more important since undergraduates may be less likely to choose an undergraduate institution for its course offerings. The primary differences are twofold: (1) we no longer see a specific disparity associated with public vs. private institutions, and (2) there is no longer a statistically significant interaction term between funding and urbanization, though in its place we see a small but statistically significant disparity on the basis of urban index alone. Note that an overall disparity between number of undergraduate QIS courses at public and private institutions is still observed ($\mu_{pub} = 0.24$, $\mu_{priv} = 0.48$); for undergraduate courses we simply cannot rule out this gap being entirely mediated via other predictors considered in the model, particularly the number of students receiving a Pell grant (see Appendix~\ref{app:correlation}).

\begin{table*}[htbp]
    \centering
    \begin{tabular}{c r c c c c c }
    \hline \hline
         \thead{Term} &  \thead{Coefficient} & \thead{$p$} & \thead{Significance} & \thead{95\% CI} & \thead{\makecell{Rate ratio}} & \thead{95\% CI}\\
        \hline
         Intercept ($a_0$) & $-2.81 \pm .25$ & $p < .001$ &  \s\s\s & [-3.32, -2.31] & -- & -- \\ 
         Dispersion ($k_2$) & $0.53 \pm .22$ & $p = .02$ & \s  & [0.09, 0.97] & -- & --
         \\ \hline
         Is\_R1 & $\ 1.76 \pm .25$ & $p < .001$ & \s\s\s & [1.26, 2.27] & 5.8 & [3.5, 9.7]  \\
         Is\_R2 & $1.12 \pm .31$ & $p < .001$ & \s\s\s & [0.50, 1.73] & 3.1 & [1.6, 5.6]  \\
         Pell\_Grant & $-0.63 \pm .13$ & $p < .001$ & \s\s\s & [-0.88,  -0.38] & $1.9^{-1}$ & [$1.5^{-1}$, $2.4^{-1}$]  \\
         Urban\_Index & $0.56 \pm .12$ & $p < .001$ & \s\s\s & [0.32, 0.81] & 1.8 & [1.4,2.2] \\ \hline \hline
         
    \end{tabular}
    \caption{Final regression results for modeling number of active undergraduate (including hybrid undergraduate/graduate) QIS courses at each institution (pseudo-$r^2$ = .17). Type II NB regression ($q=2$ in Eq.~\ref{eq:negbin-dispersion}). Significance calculated using Holm-Bonferroni adjustment for multiple statistical tests: \s \ $p_{adj} < .05$ \ \
    \s\s \ $p_{adj} < .01$ \ \
    \s\s\s \ $p_{adj} < .001$\\}
    \label{tab:results-regression-UG}

\end{table*}

\section{Distribution of degree and certificate programs}

As outlined in Sec.~\ref{sec:regression-methodology}, we also model the effect of each independent variable on the number of degree and certificate programs offered by an institution. For reference, a full list of such programs we identified can be found in Appendix~\ref{app:list}.

Our data on the distribution of degree and certificate programs exhibits no evidence of overdispersion (2-sample $t$ test, $p = 0.94$) so we adopt the simpler Poisson regression model. Results from the regression analysis are summarized in Table~\ref{tab:results-regression-degrees}. 
Once again, QIS degree programs are disproportionately found at private R1 institutions, and significant disparities are seen among public institutions as to state urbanization index and percentage of undergraduates receiving a need-based Pell grant. 

\begin{table*}[htbp]
    \centering
    \begin{tabular}{c r c c c c c }
    \hline \hline
         \thead{Term} & \thead{Coefficient} & \thead{$p$} & \thead{Significance} & \thead{95\% CI} & \thead{\makecell{Rate ratio}} & \thead{95\% CI}\\
        \hline
         Intercept ($a_0$) & $-3.80 \pm .61$ & $p < .001$ &  \s\s\s & [-5.00, -2.60] & -- & -- \\ \hline
         Is\_R1 & $\ 3.17 \pm .70$ & $p < .001$ & \s\s\s & [1.80, 4.54] & 23.8 & [6.0, 93.7]  \\
         Is\_R2 & $2.06 \pm .75$ & $p = .006$ & \s & [0.60, 3.52] & 7.8 & [1.8, 5.9]  \\
         Is\_Public & $-2.34 \pm .66$ & $p < .001$ & \s\s\s & [-3.63, -1.05] & $10.4^{-1}$ & [$2.9^{-1}$, $37.7^{-1}$] \\
         Pell\_Grant & $0.46 \pm .32$ & $p=.15$ & -- & [-0.17, 1.09] & --  \\
         Urban\_Index & $-0.01 \pm .29$ & $p=.99$ & -- & [-0.58, 0.57] & -- \\ \hline
         Public $|$ Pell & $-2.15 \pm .51$ & $p < .001$ &  \s\s\s & [-3.16, -1.14] & $8.6^{-1}$ & [$3.2^{-1}$, $23.6^{-1}$] \\
         Public $|$ Urban & $1.47 \pm .47$ & $p = .002$ & \s\s & [$0.53$, $2.40$] & $4.3$ & [$1.7$, $11.0$] \\ \hline \hline
         
    \end{tabular}
    \caption{Poisson regression results for modeling number of QIS degree and certificate programs at each institution (pseudo-$r^2$ = .30). Significance calculated using Holm-Bonferroni adjustment for multiple statistical tests: \\ \s \ $p_{adj} < .05$ \ \
    \s\s \ $p_{adj} < .01$ \ \
    \s\s\s \ $p_{adj} < .001$}
    \label{tab:results-regression-degrees}

\end{table*}

\section{Discussion and conclusions}
\label{sec:conclusions}

\subsection{Discussion}

Despite calls to emphasize equity and inclusion in the development of QIS education programs \cite{Aiello:2021,Singh:2021,Perron:2021}, our work demonstrates that clear disparities still exist in where these programs are being implemented. Both QIS coursework and degrees are disproportionately found at private R1 research institutions. These institutions tend to cater to a student body that is markedly wealthier than the average institution (see Fig.~\ref{fig:income}), resulting in an inequitable pipeline even if QIS education programs themselves are representative of the demographic makeup of the universities in which they are housed (rarely the case for STEM disciplines in practice). We also see evidence that QIS education programs appear rarer in rural states and at institutions serving higher populations of low-income students, risking the exclusion of rural and low-income populations from quantum careers.

Unlike Ref.~\cite{Cervantes:2021}, we do not find MSI status to be a statistically significant predictor of either the number of QIS courses or degrees offered by an institution. We hope this means that recent initiatives intended to promote QIS education at MSIs -- such as the IBM-HBCU Quantum Center \cite{Lee:2021} and an NSF-funded webinar at Colorado School of Mines in 2022 \cite{Mines-webinar} -- are making progress in closing this gap. A more sober assessment could be that disparities in access on the basis of MSI status as identified in Ref.~\cite{Cervantes:2021} may still be present, just primarily mediated by other institutional traits such as Carnegie classification or the percentage of low-income students. Our analysis of cross-correlation between independent variables in Appendix~\ref{app:correlation} lends credence to this latter interpretation: strong correlation ($r=0.62$) between the independent variables Is\_MSI and Pell\_Grant warrants concern about racial equality and equity in access to QIS educational programs even if the effect is primarily mediated through income rather than race, though our findings suggest that approaches sensitive to income are likely necessary to remedy these gaps.

The existence and direction of equality gaps, though saddening, may not be surprising to the reader. We draw the reader's attention not just to the \textit{existence} of equality gaps (a reality throughout much of US higher education) but to the \textit{magnitude} of the gaps we document. The effect sizes reported in Tab.~\ref{tab:results-regression}, ~\ref{tab:results-regression-UG}, and ~\ref{tab:results-regression-degrees} are, in many instances, fairly large. Per Ref.~\cite{Olivier:2017}, we interpret a rate ratio of 1.9 as a medium effect and 3.0 as a large effect. For both total QIS courses and undergraduate courses only, the number of courses students have access to at R1 research institutions was, controlling for all other variables, estimated to be a \textit{factor of 5-10 or more} greater than at non-research institutions. This same rate ratio is estimated in excess of 20 for degrees and certificates -- it is not exaggeration to say at this time that such credentials are almost entirely restricted to the tiny, elite, well-resourced (see Fig.~\ref{fig:selectivity}) minority of students attending private R1 institutions.

In short, Tab.~\ref{tab:results-regression}, ~\ref{tab:results-regression-UG}, and ~\ref{tab:results-regression-degrees} demonstrate that, at present, access to QIS education programs in the US is restricted almost exclusively to students at private R1 research institutions. Figs.~\ref{fig:income} and \ref{fig:costs-attendance} show that at least (and especially) for undergraduate students, the student population of these same private R1 institutions is highly unrepresentative of the US student body as a whole in terms of wealth. While these private R1 institutions may be the institutions with the greatest resources -- and thus the greatest ability to take on the risks associated with implementing a novel course or degree program -- the result is that, intentional or not, only an elite minority of students in US higher education have access to QIS education programs. Such a disparity, if not addressed, will invariably lead to inequities further down the quantum workforce pipeline, undermining the country's own stated policy goals as outlined in Sec.~\ref{sec:aims-scope} and risking entrenchment of a 21st century ``quantum divide'' between the haves and have-nots \cite{Gercek:2024}.

\subsection{Limitations and guidance for future research}

To our knowledge, our analysis represents the most complete cataloguing of QIS courses and degrees at US institutions attempted up to this time, and our larger $N$ enables us to build and expand upon prior work (e.g.\ Ref.~\cite{Cervantes:2021}) that has lacked the sample size to make statistical claims. However, several limitations of this earlier work remain, and further research is warranted to investigate causes of the claims we make here.

Regression analysis on observational data, as we have performed here, is affected by multicollinearity leading in particular to elevated risks of Type II error (false non-rejection of null hypothesis). 
Accordingly, the absence of a variable in the model should not be interpreted as a guarantee of equality, let alone equity, along this axis; rather, correlation between institutional factors (such as R1 status and MSI status) may simply mean that any such effects are accounted for or mediated by other independent variables in the model. We also emphasize that while our conservative approach to model-building -- only including terms if they remain statistically significant under the Holm-Bonferroni correction, and verifying this choice according to the Bayesian Information Criterion -- provides safeguards against overfitting, it may also cause otherwise significant associations to be missed. Finally, since our data represents a non-random sample of institutions, any factors that influence whether institutions appear in our search at all will affect the interpretation of our regression coefficients, an effect that we argue in Appendix~\ref{app:representative} may further inflate Type II error rates. As such, non-inclusion of variables in our models should not be taken as evidence of equal access \textit{per se}, but simply a tradeoff associated with our desire to strictly control more serious Type I error (false rejection of the null hypothesis).

Regression analysis on observational data also is inherently vulnerable to confounder and mediator biases when used for causal inference \cite{Adlakha:2023}.
As such, we caution against the use of this data to make causal claims: while we demonstrate that disparities exist along the lines of factors associated with equity gaps in the literature, we cannot infer that the institutional factors we analyze here are directly \textit{responsible} for an institution's decision to offer QIS courses or degrees.\footnote{For instance, our model is unable to rule out the possibility of student interest in QIS as a mediator or confounder -- perhaps the observed tendency of R1 institutions to offer more QIS courses is due to students who apply to R1 institutions being more interested in STEM or quantum specifically. Or perhaps students at R1 institutions are disproportionately exposed to quantum amid the research-intensive atmosphere and therefore demand that QIS courses be offered at their school.} That said, we emphasize that our findings still present equity concerns even in the absence of our ability to make causal inferences; regardless of specific causation, our findings -- combined with the observation that the institutional factors we find are predictive of QIS education programs are also correlated with relatively privileged student bodies -- demonstrate that access to QIS educational programs in the US is, for whatever reason, not equal across all groups of students.

Continued qualitative research on the experiences of underrepresented and underprivileged students in quantum remains highly important. In the absence of the ability to conduct controlled experiments assigning universities to teach QIS programs, such qualitative studies are an essential tool for understanding the root causes of such disparities so they can be addressed and developing possible solutions such as those discussed in Sec.~\ref{sec:policy}. In order to track progress in closing the gaps we identify, it will also be valuable to periodically repeat the type of quantitative analysis performed here on updated data. Future quantitative studies will likely also benefit from greater statistical power as the number of QIS courses and degrees offered continues to increase.

\subsection{Implications for educators, policymakers, and researchers}
\label{sec:policy}

As is typical for a purely quantitative study, our findings demonstrate disparities disproportionately affecting underprivileged students but provide little insight into how to address them. To provide guidance to educators and policymakers who wish to close the gaps our analysis finds, we turn to existing qualitative research and real-world experiences in quantum to explore ways these gaps may be remedied.

First, we echo Ref.~\cite{Perron:2021} in calling for investment in QIS education programs at primarily-undergraduate institutions. Likewise, policymakers ought to investigate whether there are barriers to implementing QIS education programs at institutions with high enrollment of low-income students, as well as at public institutions in rural states where investment has lagged more urbanized ones. Early results from the IBM-HBCU Quantum Center \cite{Lee:2021} have shown promise in promoting engagement with quantum careers among HBCU students; the structure of this program might well be emulated in initiatives targeting additional axes of identity. A well-rounded approach to diversity and equity in quantum, targeting rural students alongside students traditionally underrepresented along the axes of race and gender, may prove especially successful in building bipartisan support for federal investment in QIS education in an era when the concepts of diversity and equity have become politically charged.

The creation of more QIS education programs at less-resourced institutions alone, however, will not solve equality, let alone equity, gaps. Women, for example, increasingly outnumber men at US colleges and universities \cite{Belkin:2021}, yet women remain a numerical and social minority in quantum \cite{Singh:2021}. To successfully close diversity gaps, Aiello et al.\ argue that QIS education programs must focus on dismantling cultural barriers and enhancing accessibility so that all students can
succeed in quantum \cite{Aiello:2021}. Culturally-sensitive curriculum development initiatives such as QuSTEAM \cite{Porter:2022}
can be leveraged to ensure course content aligns with diverse student backgrounds and values. Outside mentorship and educational communities such as Girls in Quantum \cite{GIQ} and Qubit by Qubit \cite{QBQ} likewise must play a role in diversifying the quantum pipeline and providing underrepresented students in quantum with the social and professional communities they need to thrive. Meanwhile, Q-Turn \cite{BelenSainz:2022} and the Quantum Ethics Project \cite{Arrow:2023} exemplify the power of bottom-up organizing by marginalized quantum scientists and allies in achieving tangible change -- we call on universities and funders to support such grassroots initiatives wherever they emerge.

Additionally, as of writing, we observe that there is no CIP code specific to QIS or quantum computing degree programs. Given the rapidly-evolving QIS education landscape in the US, periodic follow-up research will be necessary to ensure these statistics remain up-to-date and to stay apprised of new developments in the field. We encourage the National Center for Education Statistics to include a quantum-specific CIP in the next update in order to streamline such research.

Finally, while our analysis specifically focuses on equal access to QIS education programs at US institutions, we also highlight the urgent need for equity not only within but across nations in building quantum education infrastructure, so that the benefits of the quantum revolution do not flow disproportionately to the Global North \cite{TenHolter:2022,DiVincenzo:2017}. Similar priority ought to be given to ensuring a diverse sampling of students and institutions as discipline-based education research (DBER) in QIS matures, to ensure both that findings are generalizable and that benefits of research-based interventions in teaching and learning also flow to students at less-resourced institutions. Coupled with the well-documented phenomenon that research subjects in PER have historically been unrepresentative of physics students as a whole \cite{Kanim:2020}, the degree of stratification we observe in student access to QIS coursework suggests that QIS education researchers ought to urgently redouble our efforts to diversify subject pools.

\begin{acknowledgments}
Special thanks to Bianca Cervantes whose work in Ref.~\cite{Cervantes:2021} laid the conceptual foundations for this investigation and to Jesse Kruse for assistance with inter-rater reliability validation. We also thank Steven Pollock for valuable feedback on the manuscript. This work is supported by the NSF GRFP and NSF
Grants No.'s 2011958 and 2012147.
\end{acknowledgments}

\FloatBarrier

\appendix

\section{More on regression methodology}
\label{app:flowchart}
This section provides a more detailed description of the Poisson and Negative Binomial (NB) regression procedure for the analysis used in this article as introduced in Sec.~\ref{sec:regression-methodology}.
\begin{enumerate}

\item A list of independent variables was developed from theory and prior work. These variables are described in Sec.~\ref{sec:vars} and summarized in Table~\ref{tab:indicators}. All binary (indicator) variables were coded as either 0 or 1; all continuous variables were normalized through a standard $z$-transformation to a mean of 0 and standard deviation of 1. We also included terms corresponding to the product of an indicator variable and a continuous variable, or two continuous variables, to study interaction effects. (Interaction terms between two indicator variables were excluded due to concerns about high structural multicollinearity -- see Appendix~\ref{app:correlation}.)

\item Standard Poisson regression was conducted -- initially including all independent variables identified in the previous step -- to obtain empirical estimates for the unknown parameters $a_0$, $\{\beta_i\}$, and $\{\gamma_{ij}\}$.

\item Terms in the Poisson model were examined one-by-one, starting with the term whose coefficient had the least statistical significance. Such terms were iteratively dropped and the regression model recomputed, except where deletion of a lower-order term in some independent variable $\theta_i$ would obscure the interpretation of a higher-order term in the same $\theta_i$. (Specifically, the term $\beta_i\theta_i$ was retained unless all such terms $\gamma_{ij}\theta_i\theta_j$ for the same independent variable $\theta_i$ had already been dropped from the model.)

\item This process of iterative deletion was repeated until all remaining regression terms met the predetermined threshold for statistical significance: $\alpha=.05$ with Holm-Bonferroni correction. (Again, the term $\beta_i\theta_i$ was always retained, regardless of statistical significance, if any term $\gamma_{ij}\theta_i\theta_j$ for the same $\theta_i$ was retained in the model.)

\item A $t$-test for overdispersion [$\sigma^2(\vec\theta) > \mu(\vec\theta)$] was performed on the final Poisson regression results. If statistically significant ($p<.05$) overdispersion was present, the analysis was rerun using both NB1 and NB2 models. In such case, the negative binomial model producing the better model fit (as determined by the Bayesian Information Criterion) was selected over the simpler, but statistically invalid, Poisson regression model.

\end{enumerate}

For convenience, this procedure is summarized graphically below in Fig.~\ref{fig:flowchart}.

\begin{figure*}[]
    \label{fig:flowchart}
    \centering
    \includegraphics[scale=0.65]{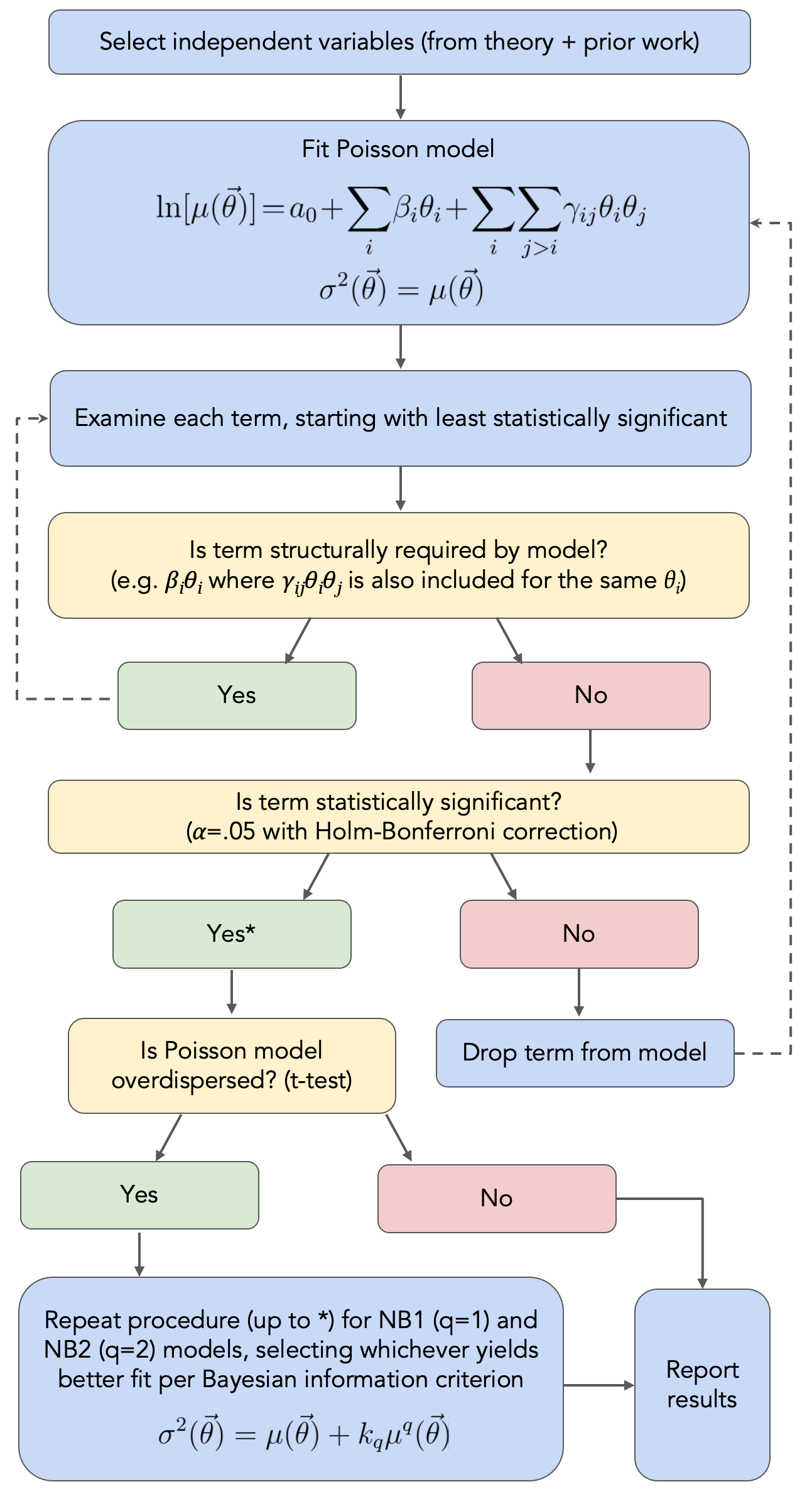}
    \caption{Flowchart illustrating the details of the regression methodology discussed in Sec.~3.4 of the main document. Here, $\vec\theta$ is a vector of independent variables representing institutional factors and $a_0$, $\beta_i$, $\gamma_{ij}$, and (for negative binomial models) $k_q$ are undetermined regression coefficients. Both the Poisson and negative binomial models model the expected mean number of counts $\mu$ and conditional variance $\sigma^2$ as a function of the independent variables $\vec\theta$. The simpler Poisson model is used unless a $t$ test indicates overdispersion.}
    \label{fig:flowchart}
\end{figure*}

\section{How representative are the institutions sampled?}
\label{app:representative}
As discussed in Sec.~\ref{sec:inst-selection}, the $N=456$ institutions we sampled were selected from those institutions that were deemed likely to have the groundwork for offering one or more QIS educational programs -- that is, already having sizable programs in one of the primary progenitor departments of physics, computer science, or electrical and/or computer engineering (ECE). While there are valid methodological reasons for this choice, one possible objection is that this sample represents a non-random sample of all institutions of higher learning in the U.S. It is certainly conceivable that disparities exist not only within our sample but between the institutions in our sample and the broader cross-section of colleges, and that these gaps could cancel one another out. We briefly discuss the implications of our non-random sample and argue that our conclusions are in general robust to sampling bias, which will tend to deflate (rather than inflate) the statistical significance and effect sizes we observe.

We compare our $N=456$ sampled institutions with $N=858$ non-sampled peer institutions in Table~\ref{tab:representativeness}. We define the sample of peer institutions as the complete list of four-year institutions in the 50 US states and the District of Columbia that meet the following criteria:
\begin{itemize}
    \item Either public or private-not-for-profit
    \item Carnegie classification as either (a) a doctoral or master's institution, (b) a baccalaureate institution with an ``arts and sciences'' or ``diverse fields'' focus, (c) a special-focus 4-year institution specializing in engineering and technology, or (d) a mixed baccalaureate/associate's-granting institution that is not primarily associate's granting,
    \item For data availability, is a Title IV, non-military institution that accepts federal need-based Pell grants, and
    \item Had an overall enrollment of at least 1000 full-time students in fall 2021. (Only 3 of the $N=456$ sampled institutions had an enrollment below 1000, so we believe this is an appropriate cutoff for comparison.)
    
\end{itemize}

As shown in Table~\ref{tab:representativeness}, our set of $N=456$ sampled institutions is indeed statistically different from the set of $N=858$ non-sampled peer institutions on every axis tested except whether the institution is an MSI. In particular, our set of sampled institutions was disproportionately public R1 and R2 institutions. Our sampled institutions were also somewhat more likely to be found in urban states and served a lower percentage of low-income students. 
We find none of these differences surprising given our methodology.

For two reasons, we believe that our results demonstrate meaningful disparities not only within the set of sampled institutions but within the set of US universities as a whole. First with the exception of the variable Is\_Public, all statistically significant variables in our regression model represented effects of the same sign as the differences observed here. (In other words, for example, R1 institutions were more likely to be among the set of sampled institutions \textit{and} more likely to, among the set of sampled institutions, offer one or more QIS courses or degrees.) Compared to a true random sample of US institutions overall, then, we would expect our regression models to understate, rather than overstate, effect sizes for most variables considered. Second, while our sampled institutions represent a minority of overall institutions of higher learning, they represent a significant majority of the overall US student body -- in other words, most students at a US 4-year institution of higher learning will find themselves represented among subset of the colleges we have searched. Table~\ref{tab:studentbody} compares the proportion of enrollment and degrees represented at sampled institutions to the broader US student body. Notice that the sampled institutions comprise approximately 2/3 of the overall enrolled full-time graduate and undergraduate student bodies, and award an overwhelming majority of degrees ($>80\%$ for bachelor's degrees, $>95\%$ for master's and doctoral degrees) for all three QIS-affiliated subjects (physics, CS, ECE) considered.

\setlength{\tabcolsep}{2pt}

\begin{table*}[!htbp]
    \centering
    \begin{tabular}{c c c c c c} 
        \hline\hline
        \thead{Variable \\ \ } &
        \thead{Sampled \\$N=456$} & \thead{Non-sampled peer \\$N=858$} & \thead{Test \\ \ } & \thead{Significance \\ \ } & \thead{Effect size \\ \ } \\ \hline
         Is\_R1 & 32\% & <1\% & Pearson $\chi^2$ & $p<.001$ & $V=0.47$ \\
         Is\_R2 & 21\% & 3\% & Pearson $\chi^2$ & $p<.001$ & $V=0.27$ \\
         Is\_Public & 63\% & 25\% & Pearson $\chi^2$ & $p<.001$ & $V=0.29$ \\
         Is\_Religious & 14\% & 33\% & Pearson $\chi^2$ & $p<.001$ & $V=0.30$\\
         Is\_MSI & 16\% & 13\% &  Pearson $\chi^2$ & $p=0.54$ & --\\ \hline
         Urban Index (unscaled) & $10.6\pm 1.0$ & $10.4\pm 1.0$ & 2-sample $t$ & $p<.001$ & $d=0.27$\\
         Pell Grant \% (unscaled) & $29\pm 12$ & $35 \pm 13$ & 2-sample $t$ & $p<.001$ & $d=-0.46$ \\
         Selectivity (\% admitted) & $68 \pm 26$ & $77 \pm 16$ & 2-sample $t$ & $p<.001$ & $d=-0.42$\\
         log(UG full-time enroll) & $8.4 \pm 1.0$ & $6.7 \pm 1.0$ &  2-sample $t$ & $p<.001$ & $d=1.4$\\
         log(Grad full-time enroll) & $6.6 \pm 1.4$ & $4.9 \pm 1.0$ & 2-sample $t$ & $p<.001$ & $d=1.3$\\
         
         \hline \hline
    \end{tabular}
    \caption{Comparison of the $N=456$ sampled institutions with $N=858$ non-sampled peer institutions.}
    \label{tab:representativeness}
\end{table*}

\begin{table*}[!htbp]
    \centering
    \begin{tabular}{l c c}
        \hline\hline
         \thead{Metric\\} & \thead{Sampled institutions ($N=456$)\\ as \% of sampled + peer ($N=1314$)}& \thead{Sampled institutions ($N=456$)\\ as \% of US 4-year ($N=2619$)} \\
         \hline
         Full-time UG enroll & 72\% & 63\% \\
         Full-time grad enroll & 80\% & 67\% \\
         \hline
         Total bachelor's awarded & 73\% & 66\% \\
         \ \ Physics & 85\% & 83\% \\
         \ \ Computer science & 92\% & 90\% \\
         \ \ ECE & 95\% & 93\% \\
         \hline
         Total master's awarded & 72\% & 61\% \\
         \ \ Physics & 98\% & 97\% \\
         \ \ Computer science & 98\% & 96\% \\
         \ \ ECE & 99\% & 97\% \\
         \hline
         Total research-based Ph.D. & 87\% & 76\% \\
         \ \ Physics & 99\% & 99\% \\
         \ \ Computer science & 99\% & 98\% \\
         \ \ ECE & 99\% & 99\% \\
         \hline \hline
         
    \end{tabular}
    \caption{Comparison of student enrollment and degrees granted at sampled institutions vs.\ all US institutions.}
    \label{tab:studentbody}
\end{table*}

\section{Cross-correlation between independent variables}
\label{app:correlation}

\begin{figure*}[!htbp]
    \centering
    \includegraphics[trim={0 0 0 0.8cm},clip,scale=0.66]{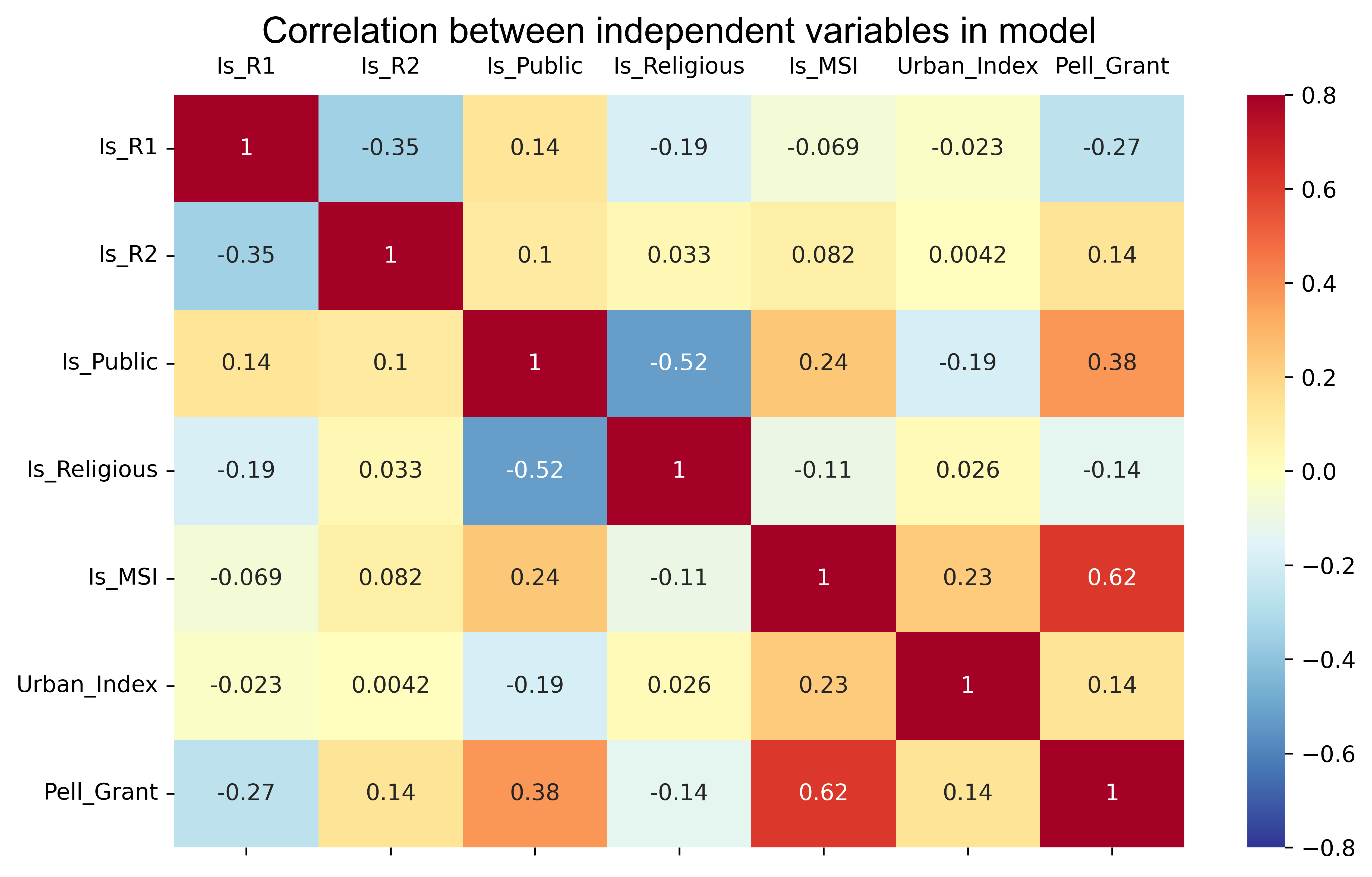}
    \caption{Correlation matrix between independent variables considered in regression model.}
    \label{fig:corr}
\end{figure*}

Multicollinearity -- cross-correlations among independent variables in a regression model -- is an important cause of Type II error. Following Cohen \cite{Cohen:1988}, we interpret a correlation of $|r|>0.3$ as moderate and $|r|>0.5$ as strong. Note that the combined effect of these correlations is generally considered within acceptable limits; all variance inflation factors for our regression models are below 6 (4 for non-intercept terms) even with inclusion of interaction terms.

\begin{itemize}
\item As expected, we see moderate-to-strong anticorrelation between mutually-exclusive indicator variables Is\_R1 and Is\_R2, and likewise between Is\_Public and Is\_Religious. These anticorrelations are structurally expected and therefore not of concern to the validity of our analysis.
\item We observe a strong positive correlation $r=0.62$ between the indicator variable Is\_MSI and the percentage of students receiving need-based Pell grants, unsurprisingly given high levels of racial income inequality in the US \cite{Akee:2019}. Strong correlation between MSI status and Pell grant percentage potentially explains the fact that in contrast to prior work \cite{Cervantes:2021}, none of our regression models found a statistically significant relationship between MSI status and the number of QIS courses or degrees offered once other variables were controlled.
\item We observe a moderate positive correlation $r=0.38$ between the indicator variable Is\_Public and the percentage of students receiving need-based Pell grants. This correlation is not large but might still be taken into account in interpreting the lack of a statistically significant correlation between Is\_Public and the number of undergraduate QIS courses offered in Table~\ref{tab:results-regression-UG}.\end{itemize}

\section{Glossary and acronyms}
\label{app:glossary}
The terms used in this article are fairly standard within the parlance of US higher education, but may be rarely encountered outside this context. To aid comprehension, we provide the following list of terms and acronyms. We anticipate this list will prove especially valuable to readers outside the US secondary education system.

\begin{itemize}

\item{\textbf{Carnegie classification of institutions of higher learning:}} A comprehensive classification of US colleges and universities provided by the nonprofit American Council on Education \cite{Carnegie}. This classification scheme is widely adopted in research concerning US institutions of higher education. Importantly, universities offering at least 20 research-based doctoral degrees annually with an annual research expenditure (2019-2020) of \$5 million or more, are classified into one of two categories:
\begin{itemize}
    \item \textit{Doctoral universities, very high research activity (R1):} if institution also meets a minimum score for research intensiveness \cite{Carnegie:Basic}.
    \item \textit{Doctoral universities, high research activity (R2):} otherwise (typically institutions with a moderate emphasis on in research, but not as high as R1s).
\end{itemize}

All remaining institutional classifications have been collapsed into a ``non-research'' designation for the purposes of this analysis, typically institutions with a primary focus on teaching.
\item{\textbf{Classification of Instructional Programs (CIP) code:}} A categorization scheme for academic programs developed by the US Department of Education. \cite{IPEDS}
\item{\textbf{Four-year college or university:}} A college or university that awards bachelor's degrees.
\item{\textbf{Integrated Postsecondary Educational Data System (IPEDS):}} A comprehensive database of US institutions of higher learning maintained by the National Center for Education Statistics within the US Department of Education. \cite{IPEDS}
\item{\textbf{Minority serving institution (MSI):}} An institution designated by NASA (the US national space agency) as serving high numbers of racial or ethnic minority students per federal law \cite{NASA:2020}. This status makes institutions eligible for additional funding and resources to support underrepresented students.

\item{\textbf{Pell grant:}} Federal program providing need-based financial aid for low-income undergraduates. The percentage of undergraduates receiving Pell grants is frequently taken as a proxy for financial need of the undergraduate student body.

\item{\textbf{Title IV:}} Federal law providing financial aid for students. A small number of institutions opt out of accepting Title IV financial aid (see footnote~\ref{fn:title-iv}); IPEDS data on such institutions is much more limited.
\item{\textbf{Undergraduate:}} Refers to students who have not yet completed a bachelor's degree.
\item{\textbf{Underrepresented minority (URM):}} Refers to individuals of a racial or ethnic minority group that is traditionally underrepresented in STEM in US higher education \cite{APS:Minority}. As used in this article, the term ``URM'' encompasses students identifying as Black or African American, Hispanic or Latino/a/x, Native American, and/or Pacific Islander.
\item{\textbf{Underrepresented minority (URM) serving:}} Refers to a minority-serving institution (MSI) subclassed as serving one or more of the underrepresented minority (URM) groups listed above. Excludes institutions that are designated exclusively as AANAPISI (Asian American and Native American Pacific Islander serving institutions -- see footnote~\ref{fn:aanapisi}). This term was adopted by our team and is not an official designation.
\end{itemize}

\FloatBarrier
\setlength{\tabcolsep}{1pt}
\renewcommand{\arraystretch}{0.7}

\begin{table*}[!p]
\vspace{-10pt}
\section{Complete list of degree and certificate programs identified}
\label{app:list}
\vspace{-10pt}
\begin{tabular}{c c c}
        \hline \hline
       \thead{Institution}  & \thead{Degree Name} & \thead{Degree Type} \\
           \hline
        U Delaware & \makecell{Quantum science \& engineering \\ Quantum science \& engineering} & \makecell{Master's, stand-alone \\ Ph.D., stand-alone} \vspace{4pt} \\
        
        U Washington Seattle & Quantum information sci \& eng & \makecell{Grad  certificate, co-enrollment} \vspace{4pt}  \\

        U California Los Angeles & Quantum sci \& tech & Master's, stand-alone \vspace{4pt}  \\

        U Texas Austin & Quantum information science & UG certificate, co-enrollment \vspace{4pt}  \\

        U Colorado Boulder & \makecell{Quantum engineering \\ Quantum information science} & \makecell{UG minor \\ Ph.D. conc., physics} \vspace{4pt}  \\

        Stony Brook U & Quantum information sci \& tech & Master's, stand-alone \vspace{4pt} \\

        U Maryland College Park & \makecell{Quantum information \\ Quantum computing} & \makecell{B.S.\ conc., CS \\ Grad certificate, professional} \vspace{4pt}  \\

        U Wisconsin Madison & Physics: quantum computing & Master's, stand-alone \vspace{4pt} \\ 

        U Pennsylvania & Photonics \& quantum tech & B.S. conc., EE \vspace{4pt} \\

        Colorado School of Mines & \makecell{Quantum eng [hardware/software tracks] \\ Quantum eng [hardware/software tracks] \\ Quantum engineering} & \makecell{Master's, stand-alone \\ Grad certificate, professional \\ UG minor} \vspace{4pt}  \\

        Harvard U & Quantum information sci and eng & Ph.D., stand-alone \vspace{4pt}  \\

        U Chicago & \makecell{Quantum sci \& eng \\ Quantum engineering} & \makecell{Ph.D., stand-alone \\ B.S.\ conc., mol.\ eng.} \vspace{4pt}\\

        U Arizona & Quantum information sci \& eng & Master's conc., optical sci. \vspace{4pt} \\

        SUNY U at Buffalo & Eng sci: Quantum sci and nanotech & Master's, stand-alone  \vspace{4pt} \\

        Rochester Inst Tech & Quantum information sci and tech & UG minor \vspace{4pt} \\

        U Pittsburgh & \makecell{Quantum computing \& quantum information \\ Physics \& quantum computing} & \makecell{UG certificate, co-enrollment \\ UG major} \vspace{4pt} \\

        George Mason U & Quantum information sci \& eng & Master's conc., physics \vspace{4pt} \\

        California Inst Tech & Quantum sci \& eng & Ph.D.\ minor \vspace{4pt}\\

        Princeton U & Quantum information \& applied physics & BS conc., ECE \vspace{4pt}\\

        Indiana U Bloomington & Quantum information sci & Master's, stand-alone \vspace{4pt}\\

        San Jose State U & Quantum computing \& information & Master's conc., EE \vspace{4pt}\\

        Emory U & Quantum information & BS conc., physics \vspace{4pt}\\

        U Colorado Denver & Quantum computing & For-credit microcredential \vspace{4pt}\\

        Stanford U & Quantum sci \& eng & BS conc., physics \vspace{4pt}\\

        U New Mexico & Quantum information sci & Ph.D.\ conc, physics \vspace{4pt}\\

        U Southern California & Quantum information sci & Master's, stand-alone \vspace{4pt} \\

        Drexel U & Quantum tech \& quantum information & Grad certificate, professional \vspace{4pt} \\

        U Massachusetts Boston & Quantum information & UG certificate [co-enroll/prof.]
        \vspace{4pt} \\

        Washington U (St. Louis) & \makecell{Quantum engineering \\ Quantum engineering} & \makecell{UG minor \\ Grad certificate, co-enrollment} \vspace{4pt} \\

        Stevens Inst Tech & \makecell{Quantum computation \\ Quantum engineering} & \makecell{Grad certificate [co-enroll/prof] \\ Master's, stand-alone} \vspace{4pt} \\

        Duke U & Quantum computing [software/hardware] & Master's conc., ECE \vspace{4pt} \\

        Harrisburg U Sci Tech & Quantum information sci & Master's conc., info sys \vspace{4pt} \\

        New York Inst Tech & Quantum informatics & B.S. conc., physics \vspace{4pt} \\

        Northwestern U & Quantum computing \& photonics & Master's conc., ECE \vspace{4pt} \\

        U Rhode Island & \makecell{Quantum computing \\ Quantum computing} & \makecell{Grad certificate, prof (online) \\ Master's, stand-alone} \vspace{4pt} \\

        St Anselm College* & Quantum information sci & UG minor \vspace{4pt} \\

        Capitol Tech U* & Quantum computing & Ph.D. (online) \vspace{4pt} \\
        
        \hline \hline\end{tabular}
    \caption{QIS degree and certificate programs at U.S. institutions as of August-September 2022. For certificate programs, co-enrollment denotes programs only available for students concurrently enrolled in another degree program; professional denotes programs intended for non-degree students that nonetheless offer university credit.\vspace{1pt}\\ \footnotesize*To verify comprehensiveness of this list, we also conducted a Google search of QIS degree programs. In doing so, we identified two programs at institutions not included in the initial catalog search. These two degree programs are excluded from the regression analysis but are included here for reference.}
    \label{tab:degrees}
    \end{table*}

\clearpage
\FloatBarrier

%

\end{document}